# Novel Codes Family for Modified Spectral-Amplitude-Coding OCDMA Systems and Performance Analysis

Mohammad Noshad, *Student Member, IEEE,* and Kambiz Jamshidi

*Abstract*—**In this paper a novel family of codes for modified spectral-amplitude-coding optical code division multiple access (SAC-OCDMA) is introduced. The proposed codes exist for more number of processing gains comparing to the previously reported codes. In the network using these codes, the number of users can be extended without any essential changes in the previous transmitters. In this study, we propose a construction method for these codes and compare their performance with previously reported codes.**

## I. Introduction

MULTI-USER interference (MUI) is the main performance degrading impairment in optical code division multiple access (OCDMA) systems. So most efforts in this area have concentrated on decreasing the bit error rate (BER) due to the MUI and consequently to increase the number of active users. In a spectrally amplitude coding OCDMA (SAC-OCDMA) system, the effect of MUI is reduced by utilizing codes with fixed in phase cross-correlation [1]. In this scheme a spectral encoding has been applied on the output of a broadband source. Some receivers have been proposed for the spectral amplitude OCDMA systems, but in all of them the system performance is degraded because of incoherency of the received light fields. Indeed in these systems the main limit on the system performance is imposed by the phase induced intensity noise (PIIN) originated from the incoherency of the light sources .The incoherent light fields that were mixed and incident upon a photodiode, causes a phase noise which lead to an intensity noise. In the square-law detection the phase of the received signal in different frequencies is not considered and therefore beating occurs among each spectral component through the square-law detection. The effect of the interference between incoherent sources on the SAC systems performance degradation is indicated in [2].

Another problem of the SAC-OCDMA systems is the limitation on the existing codes. The codes introduced for SAC-OCDMA exist for only some special integer numbers. Hence in SAC systems the trends are toward two main aims: 1) Providing codes which exist for a wide integer numbers and 2) Reducing the PIIN. A code with length $F$, weight $w$ and in-phase cross-correlation $\lambda$ is denoted by $(F, w, \lambda)$. In spectral amplitude codes the code-length is equal to the number of the frequency bins. When $\lambda = 1$, we say the code has ideal in-phase cross correlation.

In this manuscript we present a novel family of codes for modified SAC-OCDMA systems, which gives more flexibility in the selecting of the code length. These codes can be extended in length while their weight is constant and so for increasing the number of users there is no necessity to change all of the transmitter and receiver structures. In addition the construction of these codes is easy comparing to the previous codes. A simple method based on basic mathematics for designing of the codes, is offered in this paper and the structure of two-weight codes is presented using this technique. Furthermore in this paper the exact values of the mean and variance of the decision variable is calculated using combinations and then utilizing the Gaussian approximation the performance analysis of these codes is attained with good approximation. Performance of the proposed codes has been analyzed and compared with that of previously reported codes.

The rest of the paper is organized as follows. Section II is concerned on the principles of spectral amplitude OCDMA systems. In section III the code construction is studied and the structure of two-weight codes is derived. In section IV we analyze the performance of the proposed codes. In section V we give some numerical results and compare them with the simulation outcomes and finally, conclusions are drawn in section V.

## II. Principles of the Spectral Amplitude Coding

The spectral amplitude codes are the codes which spread in the wavelength domain as shown in Fig.1. Each code-word is

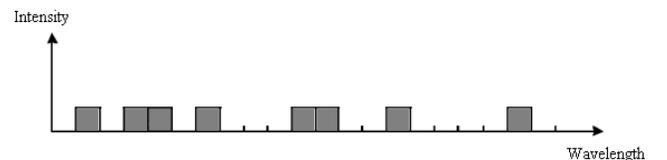

Fig.1. Scheme of spectral amplitude coding

Mohammad Noshad (e-mail: noshad@ee.sharif.edu) is with Electrical Engineering Department, Sharif University of Technology, Tehran Iran.

Kambiz Jamshidi (e-mail: jamshidi@hft-leipzig.de) is with the Department of High Frequency Technology, University of Applied Science for Telecommunication, Leipzig 04277, Germany.

TABLE I
CODE-WORDS WITH $F$=7 AND $w$=3

| Codeword | Marked Set, A($v$) |
|---|---|
| 1101000 | {1, 2, 4} |
| 0110100 | {2, 3, 5} |
| 0011010 | {3, 4, 6} |
| 0001101 | {4, 5, 7} |
| 1000110 | {1, 5, 6} |
| 0100011 | {2, 6, 7} |
| 1010001 | {1, 3, 7} |

shown by $\{c_1, c_2, ..., c_F\}$ where $c_j \in \{0,1\}$. All of the codes, used in SAC systems, are designed such that each pair of the code-words in a code family satisfy the two following properties

1) All of codes contain $w$ pulses at only $w$ wavelengths where $w$ is the weight of the codes. The other $(F - w)$ wavelengths contain no pulses. Each of the wavelengths those contain optical pulse is called *marked-wavelengths*. For a code-word with marked wavelengths of $\lambda_{k1}, \lambda_{k2}, ..., \lambda_{kw}$ we define a *marked set* and denote it by $A = \{k_1, k_2, ..., k_w\}$. Indeed the wavelength $\lambda_{kj}$ is in the marked set, if and only if $c_{kj} = 1$.

2) For each pair of codes, their *marked-sets* have only $\lambda$ common elements and the other elements of their marked set are different from each other. It means that for two code-words $a = \{a_1, a_2, ..., a_F\}$ and $b = \{b_1, b_2, ..., b_F\}$ the following relation is held

$$\sum_{k=1}^{F} a_k b_k = \lambda \qquad (1)$$

For example a sample set of these codes is shown in Table I. For these codes the code length is 7 ($F = 7$), the weight is 3 ($w = 3$) and the cross correlation is 1.

The first configuration for spectral amplitude encoding was balanced receiver (Fig.2) which is proposed by Zaccarin *et al* [1]. In this scheme the Hadamard and M-sequence codes was used for balanced detection. Most common receiver for SAC systems is differentiating receiver [3], which is depicted in Fig.3. In this system, a pulse with specified spectral distribution, $A(v)$, is sent when the data bit is "1" and nothing is sent when the data bit is "0." At the receiver side, a 1:$\alpha$ splitter is used to divide the received signal into two parts and then these parts pass through two optical filters, one with spectral pattern of $A(v)$ and the other with its complementary pattern. The spectral pattern of each user, $A_i(v)$, can be extended in the terms of the code elements of that user as follows

$$A_i(v) = \sum_{k=1}^{F} A_i^k \pi \left( \left( \frac{v - v_0}{\Delta v} \right) - \left( k - \frac{F+1}{2} \right) \right) \qquad (2)$$

where $v_0$ is the central frequency, $\Delta v$ is the bandwidth of each spectral bin as illustrated in Fig. 3. Also $A_i^k$ is the $k$th element of the $i$th user's code-word, and $\pi(x)$ is defined as

$$\pi(x) = \begin{cases} 0 & |x| > 1/2 \\ 1 & |x| \le 1/2 \end{cases}. \qquad (3)$$

If $(F, w, \lambda)$ code is used, the MUI from $(k-1)$ undesiers users at the first photodetector (PD1) is equal to $(k-1)\lambda$, and that at PD2 is equal to $\alpha(\omega - \lambda)(k-1)$. For $\alpha = \lambda/(\omega - \lambda)$, the MUI components of the branches will be equal. Therefore, after balanced photo-detection, the mean of MUI can be canceled [4]. It means that the desired user's signal can be obtained from

$$X = \sum_{i \in A_1} I(\lambda_i) - \frac{\lambda}{w - \lambda} \sum_{i \notin A_1} I(\lambda_i) \qquad (4)$$

and so the decision will be made on $X$.

Various code families have been proposed based on this receiver structure. The BIBD codes have been introduced by Zhou and *et al.* [4]. These codes are defined as $((p^{m+1} - 1/p - 1), (p^m - 1/p - 1), (p^{m-1} - 1/p - 1))$ where $p$ and $m$ are a prime and an integer numbers respectively. For the case $m = 2$ the codes have ideal cross correlation, and the code-length and code-weight are $p^2 + p + 1$ and $p + 1$ respectively for these codes.

The MQC (Modified Quadratic Congruence) and MFH (Modified Frequency Hopping) codes have been proposed in [5] and [6] respectively. Both kinds of these codes are in the form of $(q^2 + q, q + 1, 1)$. For MQC codes, $q$ is a prime number, where for MFH codes it can be a power of a prime number. So MFH codes exists for a much wider number of integers comparing to MQC codes, and hence in MFH family a code with the desired length can be chosen more freely. For both of these codes the maximum number of users is $q^2$.

Another design of OOCs with ideal in-phase cross correlation based on Projective Geometry have been appeared in [7]. These codes are $((p^{s(m+1)} - 1/p^s - 1), (p^{sm} - 1/p^s - 1), 1)$, where $p$ is a prime number. It offers larger flexibility in choosing the number of users than previously reported OOC.

A DW (Double Weight) and MDW (Modified Double Weight) codes have been introduced in [8]. For DW code family the code-weight is 2 and the code-length in terms of the maximum number of users, $N$, is as follows

TABLE II
CODES WITH FIXED IN-PHASE CROSS-CORRELATION FOR SAC SYSTEMS

| Code | $F$ | $N$ | $\omega$ | $\lambda$ |
|---|---|---|---|---|
| M-Sequence ($m$: integer) | $2^m - 1$ | $2^m - 1$ | $2^{m-1}$ | $2^{m-2}$ |
| Hadamard ($m$: integer) | $2^m$ | $2^m - 1$ | $2^{m-1}$ | $2^{m-2}$ |
| MQC ($p$: prime) | $p^2 + p$ | $p^2$ | $p + 1$ | 1 |
| MFH ($Q$: prime power) | $Q^2 + Q$ | $Q^2$ | $Q + 1$ | 1 |
| BIBD ($q$: prime power) | $q^2 + q + 1$ | $q^2 + q + 1$ | $q + 1$ | 1 |
| MDW | $3N + \frac{8}{3}\left[\sin\left(\frac{N\pi}{3}\right)\right]^2$ | $N$ | 4 | 1 |
| PMP ($M$: factor of $P-1$) | $P^2$ | $MP^2$ | $(P-1)/M$ | $\le 1$ |

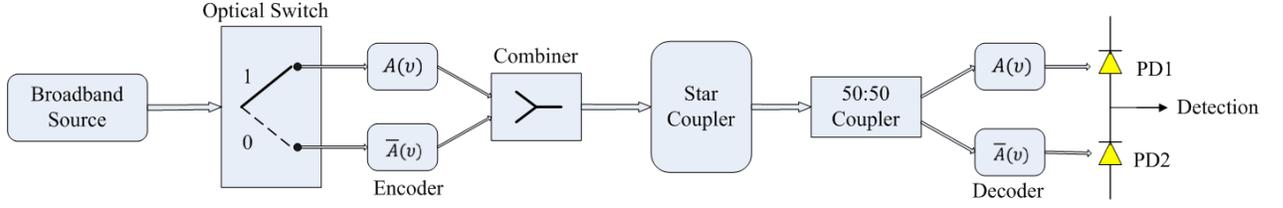

Fig.2. Balanced receiver structure.

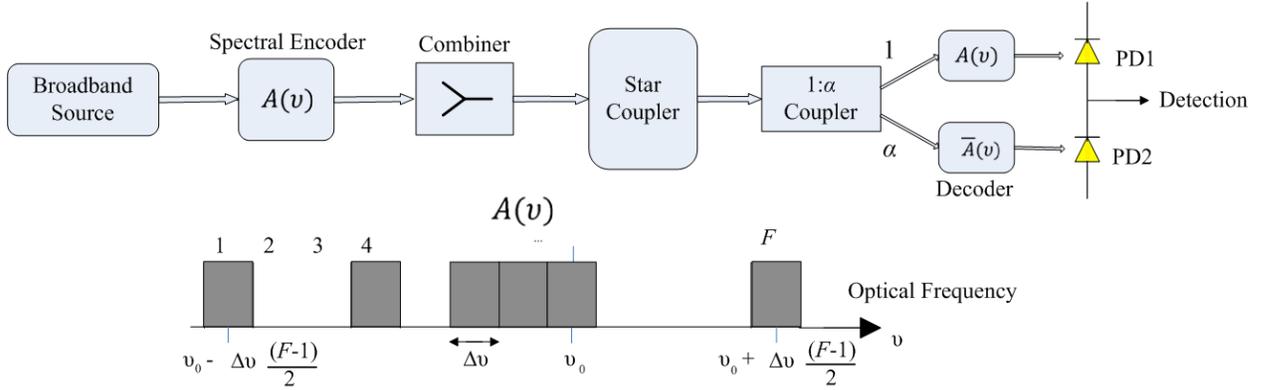

Fig.3. Differentiating receiver structure.

$$F = \begin{cases} \frac{3}{2}N & \text{for even } N \\ \frac{3}{2}N + \frac{1}{2} & \text{for odd } N \end{cases} \qquad (5)$$

MDW code is a modified DW code family that has variable weights of greater than two. For example for weight of 4 the code length is

$$F = 3N + \frac{8}{3}\left[ sin\left(\frac{N\pi}{3}\right)\right]^2 \qquad (6)$$

In this family of codes the number of users can be extended with fixed weight. So they have better performance comparing to MFH and MQC codes. In addition they exist for all natural numbers. But the code-length is much greater than the maximum number of users.

Lin et al. have been proposed the PMP (Partial modified prime) codes in [10] as

$$\left( MP^2, \frac{P-1}{M}, 1 \right) \qquad (7)$$

where $M$ is a factor of $(P-1)$. This code family has not fixed cross-correlation and $\lambda \leq 1$. Since the beating rate can be reduced by increasing the value of the dividing factor of the PMP codes, the proposed system can effectively suppress PIIN. A comparison between the different code families has been shown in Table II.

Let $\{A_i, A_j\}$ denote the cross correlation between the $i$th and $j$th code-words. So for the codes used in conventional SAC systems we have

$$\{A_i, A_j\} = \sum_{k=1}^{F} A_i^k A_j^k = \begin{cases} \lambda & 1 \leq i \neq j \leq F \\ w & i = j \end{cases} \qquad (8)$$

where $A_i^k \in \{0,1\}$ for every $k$ and $i$.

We propose a novel structure for SAC systems with novel codes, referred to as extended perfect difference (EPD) codes. In our proposed scheme (Fig.4), the two split parts of the received signal, enter into two decoders with decoding spectral pattern of $B(v)$ and $\overline{B}(v)$, where $B(v)$ contains the spectral pattern from $A(v)$ and not necessarily equal to it. In other words

$$A(v) \in B(v) \qquad (9)$$

We will show that in the proposed structure the number of users can be increased without any increase in weight and only by extending the decoding spectral pattern, i.e. $B(v)$. The decoding spectral pattern of the $i$th user, $B_i(v)$, can be expressed as

$$B_i(v) = \sum_{k=1}^{F} B_i^k \pi\left(\left(\frac{v-v_0}{\Delta v}\right) - \left(k - \frac{F+1}{2}\right)\right) \qquad (10)$$

where $B_i^k \in \{0,1\}$ for every $k$ and $i$.

Each code-word in EPD code family consists of two sets: one for transmitting the data (marked set) and the other one for detecting the received signal (detection set). We define $A_i = \{A_i^1, A_i^2, ..., A_i^F\}$ and $B_i = \{B_i^1, B_i^2, ..., B_i^F\}$ as the marked-set and detection-set of the $i$th user, respectively. The marked-set of the $i$th user and detection-set of the $j$th user satisfies the following conditions

$$\{A_i(v), B_j(v)\} = \sum_{k=1}^{F} A_i^k B_j^k = \begin{cases} \lambda & 1 \leq i \neq j \leq F \\ w & i = j \end{cases} . \qquad (11)$$

So the decoder, similar to the ordinary SAC-OCDMA systems, would be able to cancel the MUI by setting $\alpha = \lambda/(\omega - \lambda)$. Also we suppose that the proposed codes have the property that each cyclic shift of a code-word is also a code-word. Indeed by applying a cyclic shift of $(j - i)$ on the $i$th user's code-word the code-word of $j$th user is obtained. So if $A_i = \{A_i^1, A_i^2, ..., A_i^F\}$ and $B_i = \{B_i^1, B_i^2, ..., B_i^F\}$ be the marked set and detection set of the $i$th user, the marked and detection sets of the $j$th user are

$$A_j = \{A_i^1 \oplus (j-i), A_i^2 \oplus (j-i), ..., A_i^F \oplus (j-i)\}$$

$$B_j = \{B_i^1 \oplus (j-i), B_i^2 \oplus (j-i), ..., B_i^F \oplus (j-i)\}$$

where $\oplus$ is the addition module $F$.

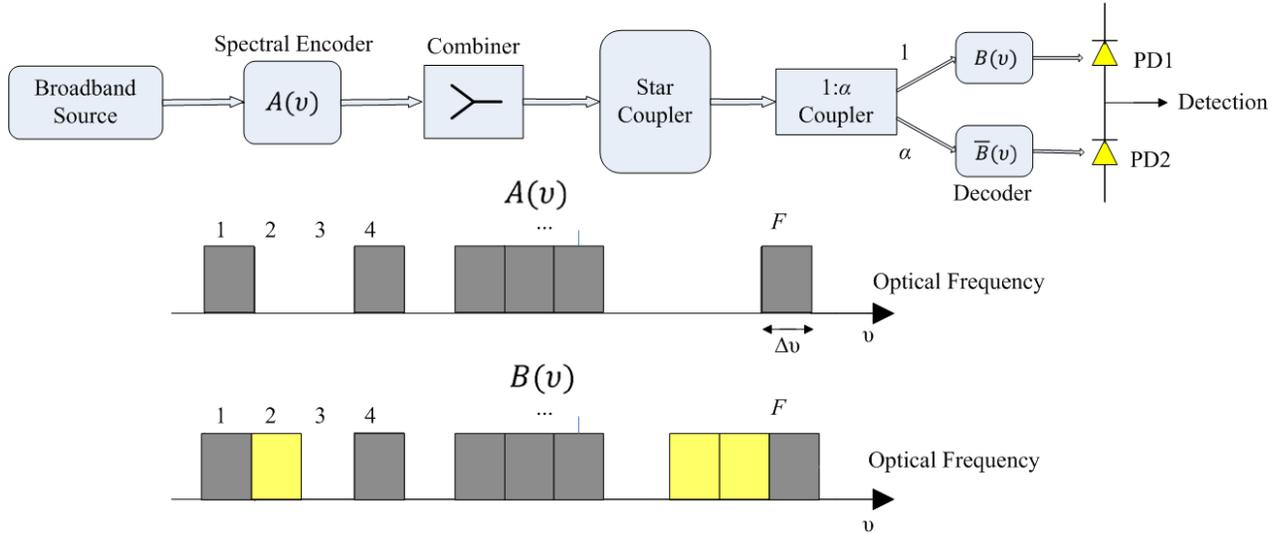

Fig.4. Structure of the proposed system.

In the remaining of the paper we use the other representation for marked and detection sets in which we only express the "1"s positions (similar to Table I).

## III. Code Construction

We denote the novel proposed code family, EPD, with the notation $(F, w, v, \lambda)$, where $F$ is the code length, $w$ is the code weight, $v$ is the detection weight, and $\lambda$ is the cross-correlation. Indeed the size of the marked set, $A$, and detection set, $B$ are $w$ and $(w + v)$ respectively. According to the cyclic shift property of these codes, the number of code-words is equal to the code length, $F$.

We first show that the code length, $F$, is equal to $w(w + v - 1) + 1$. Here we discuss the construction of the codes with ideal cross correlation ($\lambda = 1$). Without loss of generality the first user is considered as the desired user and we denote the marked and detection sets of the first user as $A_1 = \{a_1, a_2, ..., a_w\}$ and $B_1 = \{a_1, a_2, ..., a_w, a_{w+1}, ... a_{w+v}\}$ respectively. Actually $A_1$ denotes the mark positions and $B_1$ is the set that is used for detecting and decoding the received signal. For this case we consider that $a_k < a_{k+1}$ where $1 \leq k < v + w$. We construct two new sets as follows

$$\boldsymbol{T} = \{\tau_1, \tau_2, ... \tau_{w-1}\} \tag{12}$$

$$\boldsymbol{S} = \{s_1, s_2, ... s_v, s_{v+1}\} \tag{13}$$

where $\tau_i = a_{i+1} - a_i$ and $s_j = a_{w+j} - a_{w+j-1}$ for $j \leq v$ and $s_{v+1} = F + 1 - a_{w+v}$ (Fig. 5). By this definition the following relation stands for the codes

$$\left(\sum_{i=1}^{w-1} \tau_i\right) + \left(\sum_{i=1}^{v+1} s_i\right) = F. \tag{14}$$

Because of the cyclic property of these codes, $\boldsymbol{T}$ and $\boldsymbol{S}$ of each code-word are equal to $\boldsymbol{T}$ and $\boldsymbol{S}$ of the first code-word, respectively. The marked-sets of the other codes are the cyclic shifts of that of the first user, $A_1$, and because of the ideal cross-correlation property of these codes, the cyclic shifts of the $A_1$ must have exactly one common element with the detection-set of the first user, $B_1$. So for each cyclic shift of $A_1$, only one element of the marked-set coincide with one of the $(w + v - 1)$ elements of the $B_1$. Considering all the $w$ elements of the marked-set, there can exist only $w(w + v - 1)$ cyclic shifts of $A_1$. Thus the total number of the code-words will be $w(w + v - 1) + 1$, which must be equal to the code-length, $F$. This means that $F = w(w + v - 1) + 1$.

Now we construct the extended sets of $\boldsymbol{T}$ and $\boldsymbol{S}$ as follows

$$\boldsymbol{E_T} = \left\{\left(\sum_{i=k}^{l} \tau_i\right), \left(F - \sum_{i=k}^{l} \tau_i\right) \mid 1 \leq k \leq l \leq w-1\right\} \tag{15}$$

$$\boldsymbol{E_S} = \left\{\left(\sum_{i=k}^{l} s_i\right) \mid 1 \leq k \leq l \leq v+1\right\} \tag{16}$$

$$\boldsymbol{E_{ST}} = \left\{\left(\sum_{i=k}^{w-1} \tau_i\right) + \left(\sum_{i=1}^{l} s_i\right) \mid 1 \leq k \;, l \leq v\right\}. \tag{17}$$

For fulfilling the ideal cross-correlation property, the distance between the two elements of $A_k$ ($k \neq 1$) shouldn't be equal to the distance of different pair of elements in $B_1$, because in this case two elements of $A_k$ ($k \neq 1$) coincide with the two elements of $B_1$ and the ideal cross-correlation condition will be violated. So two conditions can be derived regarding these statements as follows

1) All of the elements in $E_T$ must be distinct, and

2) $E_T$ shouldn't have any common element with neither of $E_S$ nor $E_{ST}$.

From the first condition it's inferred that $E_T$ has $w(w-1)$ components, i.e. $n(E_T) = w(w-1)$.

From the Appendix A we know that

$$s_1 = s_2 = \cdots = s_v = s_{v+1} = s. \qquad (18)$$

By eliminating the repeated elements in $E_S$ and $E_{ST}$, they can be written as follows

$$E_S = \{ks|\ 1 \le k \le v+1\} \qquad (19)$$

$$E_{ST} = \left\{\left(\sum_{i=k}^{w-1} \tau_i\right) + ls|\ 1 \le k\ , l \le v\right\}. \qquad (20)$$

We define a subset of $E_S$, $E_{S1}$, as follows

$$E_{S1} = \{ks|\ 1 \le k \le v\}.$$

Now consider that $E_{ST}$ has a common element with the $E_{S1}$. In this case we should have the following relation

$$\left(\sum_{i=k}^{w-1} \tau_i\right) + ls = js \Rightarrow \sum_{i=k}^{w-1} \tau_i = (j-l)s$$

so the second condition will be violated. Similarly if $E_{ST}$ has two repetitive elements then we have

$$\left(\sum_{i=k}^{w-1} \tau_i\right) + ls = \left(\sum_{i=j}^{w-1} \tau_i\right) + ms \Rightarrow \sum_{i=k}^{j-1} \tau_i = (m-l)s$$

and this is also the violation of the second condition.

Considering this we can deduce that all the elements of $E_{ST}$ are distinct and it shouldn't have any common element with neither of $E_T$ and $E_{S1}$. Thus the number of the elements in $E_{ST}$ is $v(w-1)$, i.e. $n(E_{ST}) = v(w-1)$.

Using (14) we can obtain the first condition on the positions of the weighted bits as

$$\left(\sum_{i=1}^{w-1} \tau_i\right) = F - s(v+1)$$

$$= (w-1)^2 + (v+1)(w-s). \qquad (21)$$

On the other hand $E_T$, $E_{S1}$ and $E_{ST}$ are completely distinct sets. In other words they have no common elements. Therefore the total number of the disparate elements in all of these three sets must be equal to the sum of the sizes of these three sets. So we have

$$n(E_{ST} + E_T + E_{S1}) = n(E_{ST}) + n(E_T) + n(E_{S1}). \qquad (22)$$

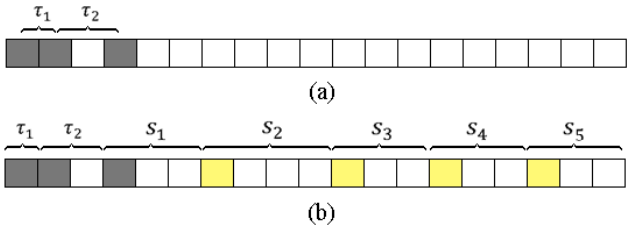

Fig.5. (a) Marked-set and (b) Detection-set of a user.

So their total distinct elements is obtained as

$$n(E_{ST} + E_T + E_{S1}) = v(w-1) + w(w-1) + v$$

$$= w(w+v-1) = F-1. \qquad (23)$$

Furthermore all the elements of these three sets are less than $F$. So the elements of these three sets cover all the integer numbers less than $F$. So the summation of their elements should be equal to the sum of all integer numbers from 1 to $(F-1)$. So

$$\sum_{x_i \in E_T} x_i + \sum_{x_i \in E_{S1}} x_i + \sum_{x_i \in E_{ST}} x_i = 1 + 2 + \cdots + F - 1$$

$$= \frac{F(F-1)}{2}. \qquad (24)$$

From (12), (13) and (14) the sum of elements of each set can be expressed as

$$\sum_{x_i \in E_T} x_i = \sum_{k,l=1}^{w-1} \left(\sum_{j=k}^{l} \tau_j\right) + \sum_{k,l=1}^{w-1} \left(F - \sum_{j=k}^{l} \tau_j\right)$$

$$= \frac{w(w-1)F}{2} \qquad (25)$$

$$\sum_{x_i \in E_{S1}} x_i = \sum_{k=1}^{v} \left(\sum_{j=1}^{k} s_j\right) = \sum_{j=1}^{v} (v-j+1)s_j \qquad (26)$$

$$\sum_{x_i \in E_{ST}} x_i = \sum_{k=1}^{v} \sum_{l=1}^{w-1} \left(\sum_{j=l}^{w-1} \tau_j + \sum_{j=1}^{k} s_j\right)$$

$$= v \sum_{j=l}^{w-1} j\tau_j + (w-1) \sum_{j=l}^{v} (v-j+1)s_j \qquad (27)$$

By substituting (25), (26) and (27) in (24), we obtain

$$\frac{w(w-1)F}{2} + v \sum_{j=1}^{w-1} j\tau_j + w \sum_{j=1}^{v} (v-j+1)s_j$$

$$= \frac{F(F-1)}{2} \qquad (28)$$

and by applying (18) the following result is obtained

$$\sum_{j=1}^{w-1} j\tau_j + sw(v+1)/2 = wF/2 \qquad (29)$$

Subtracting the left hand side of (21) from that of (29), the second condition is obtained as follows

$$2\sum_{j=1}^{w-2} j\tau_{j+1} = (w-2)\left[\ (w-1)^2 + (v+1)(w-s)\ \right] \qquad (30)$$

Regarding the two conditions, (21) and (30), for a certain value of $s$ there could be a code-set. For example for the code-family with weight two, only the first criterion suffices for designing the codes. This criterion is

$$w = 2 \Rightarrow \tau_1 = 1 + (v+1)(2-s)$$

TABLE III
MARKED SETS AND DETECTION SETS OF EPD CODES
FOR $F = 7, w = 2, v = 2$ AND $s = 1$

| Marked sets | Detection sets |
|---|---|
| (1, 5) | (1, 5, 6, 7) |
| (2, 6) | (2, 6, 7, 1) |
| (3, 7) | (3, 7, 1, 2) |
| (4, 1) | (4, 1, 2, 3) |
| (5, 2) | (5, 2, 3, 4) |
| (6, 3) | (6, 3, 4, 5) |
| (7, 4) | (7, 4, 5, 6) |

TABLE IV
MARKED SETS AND DETECTION SETS OF EPD CODES
FOR $F = 7, w = 2, v = 2$ AND $s = 2$

| Marked sets | Detection sets |
|---|---|
| (1, 2) | (1, 2, 4, 6) |
| (2, 3) | (2, 3, 5, 7) |
| (3, 4) | (3, 4, 6, 1) |
| (4, 5) | (4, 5, 7, 2) |
| (5, 6) | (5, 6, 1, 3) |
| (6, 7) | (6, 7, 2, 4) |
| (7, 1) | (7, 1, 3, 5) |

$$\Rightarrow \tau_1 = \begin{cases} 1 \quad ; & s = 2 \\ v + 2; & s = 1 \end{cases}. \tag{31}$$

Hence the detection and marked sets of the code-words are as follows

$s = 1$:

$$\begin{cases} A_j = \left\{ j, j + \dfrac{(F+1)}{2} \right\} \ (mod \ F) \\ B_j = \{ j, j + \dfrac{F+1}{2}, j + \dfrac{F+3}{2}, \dots, j + F - 1 \} (mod \ F) \end{cases} \tag{32}$$

$s = 2$:

$$\begin{cases} A_j = \{ j, j + 1 \} \ (mod \ F) \\ B_j = \{ j, j + 1, j + 3, j + 5, \dots, j + F - 2 \} \ (mod \ F) \end{cases} \tag{33}$$

where $F = 2(v + 1) + 1$. The examples of these codes are listed in Tables III and IV. One can observe that the code-words in (32) and (33) satisfy (9) and the cross correlation condition in (11). As can be seen from (32) and (33), in the proposed codes increasing the number of users by 2 can be done by changing the transmitter's wavelengths of only one existing user and modifying the detection sets of all users.

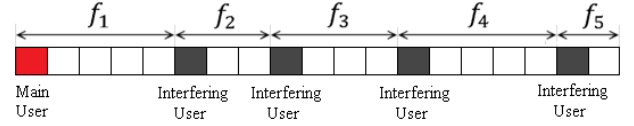

Fig. 6. Distance between the "one" transmitting users.

For the codes with weight three, according to (21) and (30), we have

$$w = 3 \Rightarrow \begin{cases} \tau_1 + \tau_2 = 4 + (v + 1)(3 - s) \\ 2\tau_2 = 4 + (v + 1)(3 - s) \end{cases} \Rightarrow \tau_1 = \tau_2$$

which can't be true. Similarly for the codes with weight four

$$w = 4 \Rightarrow \begin{cases} \tau_1 + \tau_2 + \tau_3 = 9 + (v + 1)(4 - s) \\ 2(\tau_2 + 2\tau_3) = 2[9 + (v + 1)(4 - s)] \end{cases}$$

$$\Rightarrow \tau_2 + 2\tau_3 = \tau_1 + \tau_2 + \tau_3 \Rightarrow \tau_1 = \tau_3.$$

This is in contradiction with the aforementioned conditions. Thus there are no codes neither with weight three nor four. So EPD codes exist for specific code-weights.

## IV. PERFORMANCE ANALYSIS

In this section we obtain the performance of our proposed codes. We use combinational methods in all these performance analysis. For the analysis of the proposed codes we obtain the beat noise of the receiver considering the noise caused by the incoherency of the sources. Also in analyzing the performance of the system, synchronization between different users has been considered. This consideration is taken only for mathematical sake and is not necessary for the system operation and the system can be utilized properly without any synchronization between users. Here we use a combinational method to calculate the performance of the system. We first calculate the pdf (probability density function) of the decision variable defined in (4) for the codes with weight two.

So consider an optional code for which $\tau$ and $F$ are *coprime* (relatively prime). We rearrange the bits of the code in another order to show that its performance is the same as the code with $\tau = 1$. Map the bit $k\tau + 1(mod \ F)$ to the $k$th bit and so each two bits with a distance of $\tau$ (contains cycled end case) be adjacent after the mapping. It's clear that this is a *one to one* transformation because the GCD (Greatest Common Divisor) of $\tau$ and $F$ is 1. Thus we concentrate on the code family with $\tau = 1$, which is indicated in (33).

The pdf of the decision variable $X$ can be expressed by using the Bayesian rule as

$$f_X(x|b) = \sum_{m=0}^{N-1} f_X(x|m, b)P(m) \tag{34}$$

where $N$ is the number of active users and $P(m)$ is the probability that $m$ interfering users send mark in the transmitting bit period, which for the equal probable transmitting equals

$$P(m) = \binom{N-1}{m} \left(\frac{1}{2}\right)^{N-1} \tag{35}$$

$f_X(x|m)$ in (34) has two different forms for the data bits "one" and "zero" and is shown in (36) and (37) respectively.

$$f_X(x|m, b=1) = \sum_{j=0}^{m+1} P_1(n=j|m) f_X(x|n=j, b=1) \quad (36)$$

$$f_X(x|m, b=0) = \sum_{j=0}^{m-1} P_0(n=j|m) f_X(x|n=j, b=0) \quad (37)$$

where in the both of these equations $P_i(n|m)$ is the probability of $n$ interferences conditioned on $m$, and $f_X(x|n, b)$ is the pdf of $x$ conditioned on $n$ interferences.

We consider the first user as the main user and suppose that the users $k_1, k_2, \ldots, k_m$ transmit "one" and the remaining users transmit "zero". For calculating $P_1(n|m)$ first we define a new set $f = \{f_1, f_2, \ldots, f_{m+1}\}$, where $f_i$ is defined as the distance between the $k_{i-1}$th and $k_i$th users, and is equal to $(k_i - k_{i-1})$, where $k_0 = 1$ and $k_{m+1} = F + 1$. For example for $k_1 = 6, k_2 = 9, k_3 = 13, k_4 = 18$, and $F = 19$ $f$ is $\{5,3,4,5,2\}$, as indicated in Fig. 6. So $f_i$s are integer numbers greater than or equal to 1 and the following relation is held between them.

$$f_1 + f_2 + \cdots + f_{m+1} = F. \quad (38)$$

For the codes defined in (33), the occurrence of an interference means that the distance between two successive users is 1. So the probability of occurrence of $n$ interferences can be obtained by calculating the number of solutions of (38) in which $n$ of $f_i$s is 1 and the others are greater than one. By setting $f_{k_1} = f_{k_2} = \cdots = f_{k_n} = 1$, (38) becomes as

$$f_{k_{n+1}} + f_{k_{n+2}} + \cdots + f_{k_{m+1}} = F - n \quad (39)$$

such that $f_{k_i} > 1$ for $i = j + 1, \ldots, m + 1$. We define new variables, $u_i = f_{k_{j+i}} - 1$ for solving (39). By substituting these variables (39) becomes as

$$u_1 + \cdots + u_{m-n+1} = F - m - 1 \quad (40)$$

So $P_1(n|m)$ is obtained as follows

$$P_1(n|m) =$$
$$\frac{\binom{m+1}{n} \times (\text{Number of solutions of (40)})}{\text{Number of solutions of (38)}}. \quad (41)$$

So (41) can be rewritten in the form of

$$P_1(n|m) = \frac{\binom{m+1}{n}\binom{F-m-2}{m-n}}{\binom{F-1}{m}}. \quad (42)$$

Thus (34) becomes in the form of (43), for mark bit.

$$f_X(x|b=1) = \frac{(N-1)!}{(F-1)!} \left(\frac{1}{2}\right)^{N-1}$$
$$\times \sum_{m=0}^{N-1} \sum_{j=0}^{m+1} \binom{m+1}{j}\binom{F-m-2}{m-j}$$
$$\times \frac{(F-m-1)!}{(N-m-1)!} f_X(x|n=j, b=1). \quad (43)$$

The pdf of $x$ in the case of transmitting "zero" can be obtained in a similar manner. In this case, the main user does not send any signal and therefore $f_1 = 1$ and $f_{m+1} = 1$ doesn't cause any interference. So we don't apply the change of variables for neither of them in (39) and it becomes as follows

$$r_1 + u_1 + \cdots + u_{m-j-1} + r_{m+1} = F - m + 1 \quad (44)$$

and hence $P_0(n|m)$ is obtained as

$$P_0(n|m) = \frac{\binom{m-1}{n} \times (\text{Number of solutions of (44)})}{\text{Number of solutions of (38)}}$$
$$= \frac{\binom{m-1}{n}\binom{F-m}{m-n}}{\binom{F-1}{m}}. \quad (45)$$

Thus for a "zero" bit (34) becomes in the following form

$$f_X(x|b=0) = \frac{(N-1)!}{(F-1)!} \left(\frac{1}{2}\right)^{N-1}$$
$$\times \sum_{m=0}^{N-1} \sum_{j=0}^{m} \binom{m-1}{j}\binom{F-m}{m-j}$$
$$\times \frac{(F-m-1)!}{(N-m-1)!} f_X(x|n=j, b=0). \quad (46)$$

By the Gaussian approximation

$$f_X(x|n, b=1) = \frac{1}{\sqrt{2\pi n\sigma_0^2}} exp\left[\frac{(x-2I_0)^2}{n\sigma_0^2}\right] \quad (47)$$

$$f_X(x|n, b=0) = \frac{1}{\sqrt{2\pi n\sigma_0^2}} exp\left[\frac{x^2}{n\sigma_0^2}\right] \quad (48)$$

where $\sigma_0^2$ and $I_0$ are the variance and mean of the photo detector current due to the interference of the two optical fields. According to [9], $\sigma_0^2$ and $I_0$ are defined as follows

$$\sigma_0^2 = 2\frac{B_e}{\Delta v}\left(\frac{RP_{sr}}{F}\right)^2 \quad (49)$$

TABLE V
PARAMETERS USED IN THE SIMULATION AND NUMERICAL CALCULATIONS

| Quantum Efficiency of the Photo-detector | $\eta = 0.6$ |
|---|---|
| Optical Band-Width of a Spectral Bin | $\Delta v = 28\ GHz$ |
| Electrical Band-Width of the Receiver | $B_e = 80\ MHz$ |
| Central Wavelength | $\lambda_0 = 1.55\ \mu m$ |
| Bit Rate | $R_b = 155\ Mbps$ |
| Power of Each User at the Receiver | $P_0 = -10dbm$ |
| Receiver Noise Temperature | $T_r = 300\ K$ |
| Receiver Load Resistor | $R_L = 1000\ \Omega$ |

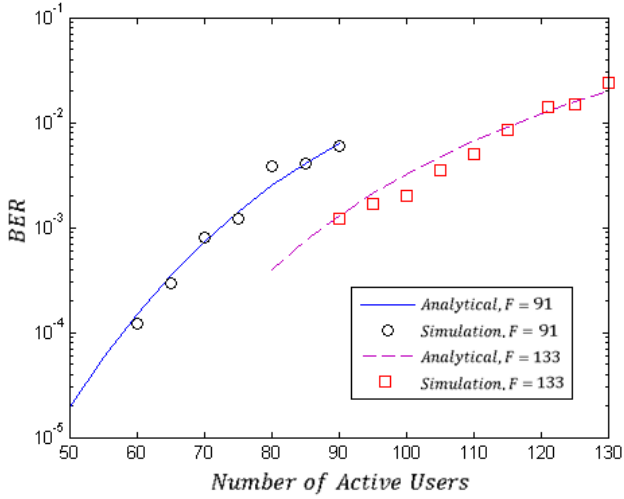

Fig. 7. Simulation and analytical results of BER for the proposed codes.

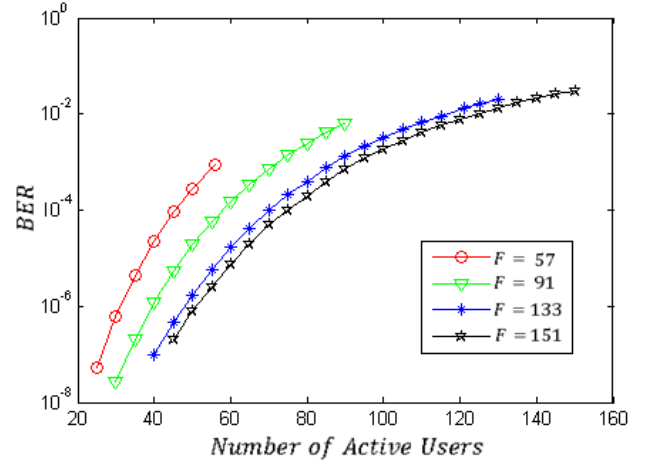

Fig. 8. Analytical results of BER versus the number of active users for different code-lengths.

$$I_0 = \frac{RP_{sr}}{F} \quad (50)$$

where $R$ is the responsivity of the photo-detector, $P_{sr}$ is the total transmitted power by the source, $B_e$ is electrical bandwidth, $\Delta v$ is optical bandwidth of a spectral bin which is equal to $B_o/F$ and $B_o$ is the total optical bandwidth. So the probability of error can be obtained as follows

$$Pe = \int_{-\infty}^{Th} f_X(x|n, b=1) \, dx + \int_{Th}^{\infty} f_X(x|n, b=0) \, dx \quad (51)$$

where $Th$ is the optimum decision threshold.

## V. NUMERICAL RESULTS AND SIMULATION

In this section the results of the simulations are compared with the results of the previous section. Furthermore, Performance analysis of the system using proposed codes will be made and compared with that of MQC and MFH codes. Phase induced intensity noise (PIIN) has been considered as the main performance degrading impairment of SAC-OCDMA systems [2]. In the first part, performance of the novel proposed codes in the section IV is evaluated in order to see the accuracy of these equations comparing to the simulation results. The simulations have been accomplished using Mont-Carlo method for an equal probable bit transmission. The iteration of runs continues until at least 10 errors occur at the receiver side. All of the phases of optical fields in all of the spectral bins for all users supposed to be independent and have non-stationary zero-mean Gaussian distribution as indicated in [11]. Furthermore the transmitted power for all of the users has been considered to be equal. The loss and dispersion of the optical path have been neglected.

Fig. 7 shows the bit error rate versus the number of active users for the proposed codes with weight two and code-length 133 and 91 for both simulation and analytical results. In this case the shot noise and thermal noise have been neglected. These results are obtained for a system with the parameters listed in Table V. One can observe the good matching between the simulation and analytical results. The bit error rates for four different code-lengths are depicted in Fig. 8 versus the number of active users. As can be seen, performance of the system improves as code length increases.

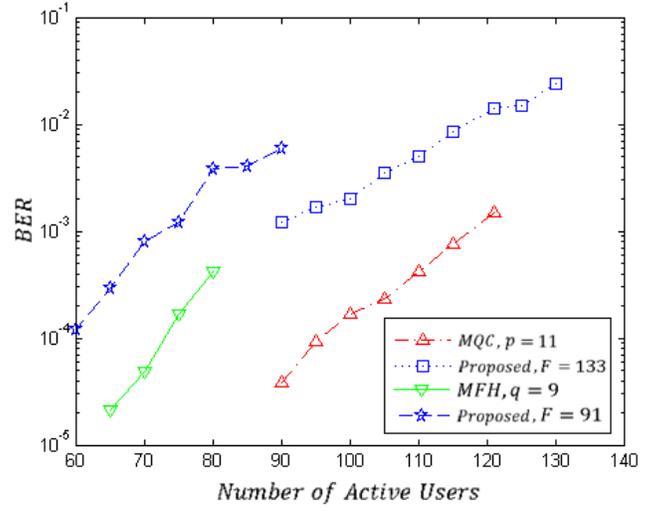

Fig. 9. Simulation results for MQC ($p$=11), MFH ($q$=9) and proposed ($F$=91, $F$=133) codes.

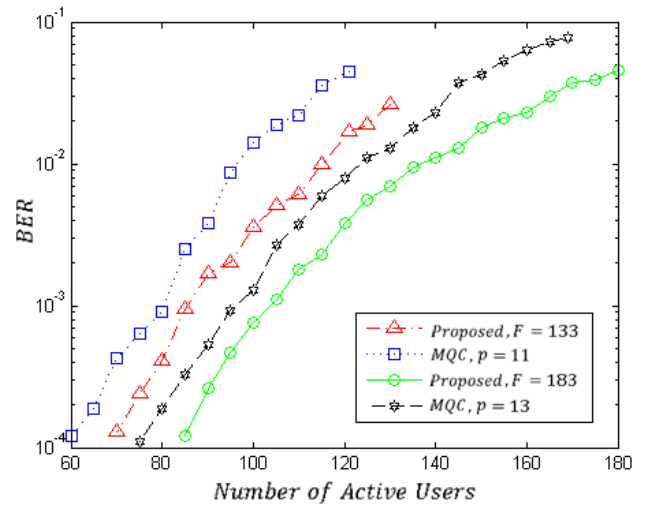

Fig. 10. Simulation results for MQC ($p$=11 and $p$=13) and proposed ($F$=133 and $F$=183) codes.

Comparison of the MQC ($p$=11) and MFH ($q$=9) codes with proposed ($F$=91, $F$=133) codes is illustrated in Fig. 9. For the simulating of the MQC codes we use the codes family with $\alpha$=1, $\beta$=0 and b=2. As can be seen in this figure, the performance of the proposed codes is somehow worse than the previously reported codes.

For the traditional codes the output of the laser source is filtered by the signature spectral pattern. So a part of the laser's power is wasted in this filtering and a small part of the laser source is transmitted. But for the aforementioned novel proposed codes with weight two, the laser power can be transmitted without any filtering in the two adjacent spectral bins. Thus most part of the laser source can be used in the data transmission and the transmitted data can dominate the shot noise. The simulation results in Fig. 10 reveal the superiority of the proposed codes when thermal noise and shot noise are included. In this figure the total power of each user considered to be equal for all codes.

## VI. CONCLUSION

In this paper a novel structure for SAC-OCDMA receivers has been introduced. This proposed system differs with the ordinary systems in the receiver structure and the code length can be extended to vast numbers without any increase in the code-weight. Indeed it has been shown that the codes with weight two exist for code-lengths in the form of $(2k + 1)$, where $k$ is an integer number greater than zero. Also it has been shown that the codes do not exist for $w = 3$ and $w = 4$, and so it can be concluded that there is a limitation in selecting code-weight. But instead there is not much limitation in selecting code-length.

We offer a construction method for these codes using a combinational methods and focus on the codes with weight two. The performance of these codes has been presented by a good approximation and the simulation results show a good match with the analytical results. The other benefit of the proposed codes is their simple structure and convenience in the implementation. A comparison between the simulation results of the proposed codes, MQC codes and MFH codes show a little bit worse performance for our proposed codes. Instead these codes can use the approximately all of the laser source power for transmitting the data bit, because its weighted frequencies are adjacent. So these codes surpass the traditional codes assuming the shot noise.

## APPENDIX A

Using induction we prove that for every element in $\boldsymbol{S}$ there exists an integer number, $k_j$, such that

$$s_j = s_1 + \sum_{i=k_j}^{w-1} \tau_i \quad ; \ 2 \leq j \leq v+1 \tag{A.1}$$

If we shift the code word to the right by the value of $\sum_{i=x}^{y} s_i$ then one of the marked bits, $a_m$, must coincides on the one of the detection bits, $a_n$. But if $1 \leq n \leq w$ then it inferred that

$$\sum_{i=x}^{y} s_i = \sum_{i=n}^{m-1} \tau_i \tag{A.2}$$

and it violates the second condition. Thus $w + 1 \leq n \leq v$ and therefore

$$\sum_{i=x}^{y} s_i = a_n - a_m = \sum_{i=n}^{w-1} \tau_i + \sum_{i=1}^{l} s_i \tag{A.3}$$

where $l = m - w < y$. Hence for $x = y = 2$ we have

$$s_2 = s_1 + \sum_{i=n}^{w-1} \tau_i \tag{A.4}$$

So (A.1) is true for $j = 2$ and it is the base of the induction. Now consider that

$$s_j = s_1 + \sum_{i=h}^{w-1} \tau_i \tag{A.5}$$

is true for $1 \leq j \leq k$. Here $h$ is an integer number. From (A.3) we know that

$$\sum_{i=2}^{k} s_i = \sum_{i=n}^{w-1} \tau_i + \sum_{i=1}^{l} s_i \tag{A.6}$$

So we have

$$\sum_{i=l+1}^{k+1} s_i = \sum_{i=n}^{w-1} \tau_i + s_1 \tag{A.7}$$

So if $l < k$, then according to (A.5) there exists an integer number, $u$, such that for $s_{l+1}$ we have

$$s_{l+1} = s_1 + \sum_{i=u}^{w-1} \tau_i \tag{A.8}$$

and thus

$$\sum_{i=l+2}^{k+1} s_i + \left( s_1 + \sum_{i=u}^{w-1} \tau_i \right) = \sum_{i=n}^{w-1} \tau_i + s_1 \tag{A.9}$$

By simplifying (A.9) we get

$$\sum_{i=l+2}^{k+1} s_i = \sum_{i=n}^{u-1} \tau_i \tag{A.10}$$

and it is a contradiction with the second condition. So in (A.7) $k$ can't be greater than $l$ and they must be equal ($l = k$). Therefore

$$s_{k+1} = \sum_{i=n}^{w-1} \tau_i + s_1. \tag{A.11}$$

Hence (A.1) is valid for all of the numbers between 2 and $v$.

Similarly we can prove that for every $j$ there exists an integer number $l_j$ such that

$$s_j = s_{v+1} + \sum_{i=1}^{l_j} \tau_i, \ \ 1 \leq j \leq v \quad\quad (A.12)$$

From (A.1) and (A.12) for $s_1$ and $s_{v+1}$ we have

$$s_{v+1} = s_1 + \sum_{i=k_j}^{w-1} \tau_i \quad\quad (A.13)$$

$$s_1 = s_{v+1} + \sum_{i=1}^{l_j} \tau_i \quad\quad (A.14)$$

and therefore

$$s_1 = s_{v+1} \quad\quad (A.15)$$

From (A.1) and (A.12) for $1 < j < v + 1$ we can conclude

$$s_{v+1} + \sum_{i=1}^{l_j} \tau_i = s_1 + \sum_{i=k_j}^{w-1} \tau_i \quad\quad (A.16)$$

and applying (A.15) results

$$\sum_{i=1}^{l_j} \tau_i = \sum_{i=k_j}^{w-1} \tau_i \quad\quad (A.17)$$

But it is in an obvious contradiction with first condition. So (A.1) and (A.12) must be in the following form

$$s_j = s_1 = s_{v+1}, \quad\quad 2 \leq j \leq v \quad\quad (A.18)$$

## References

[1] D. Zaccarin and M. Kavehrad, "An optical CDMA system based on spectral encoding of LED," *IEEE Photon. Technol. Lett.*, vol. 4, pp. 479–482, Apr. 1993.

[2] E. D. J. Smith, R. J. Blaikie, and D. P. Taylor, "Performance enhancement of spectral-amplitude-coding optical CDMA using pulse-position modulation," *IEEE Trans. Commun.*, vol. 46, no. 9, pp. 1176–1185, Sep. 1998.

[3] E. D. J. Smith, P. T. Gough, and D. P. Taylor, "Noise limits of optical spectral-encoding CDMA systems," *Electron. Lett.*, vol. 31, no. 17, pp. 1469–1470, Aug. 1995.

[4] X. Zhou, H. M. H. Shalaby, C. Lu, and T. Cheng, "Code for spectral amplitude coding optical CDMA systems," *Electron. Lett.*, vol. 36, pp. 728–729, Apr. 2000.

[5] Zou Wei, H. Ghafouri-Shiraz, and H. M. M. Shalaby, "New Code Families for Fiber-Bragg-Grating-Based Spectral-Amplitude-Coding Optical CDMA Systems," *IEEE Photon. Technol. Lett.*, vol. 13, pp. 890–892, Aug. 2001.

[6] Zou Wei and H. Ghafouri-Shiraz, "Proposal of a Novel Code for Spectral Amplitude-Coding Optical CDMA Systems" *IEEE Photon. Technol. Lett.*, vol. 14, no. 3, Mar. 2002.

[7] I. B. Djordjevic and B. Vasic "Unipolar Codes for Spectral-Amplitude-Coding Optical CDMA Systems Based on Projective Geometries", *IEEE Photon. Technol. Lett.*, vol. 15, no. 9, Sep. 2003.

[8] S. A. Aljunid, M. Ismail, A. R. Ramli, , B. M. Ali and M. K. Abdullah, "A New family of optical code sequences for spectral-amplitude-coding optical CDMA systems," *IEEE Photon. Technol. Lett.*, vol. 16, pp. 2383–2385, Oct. 2004.

[9] M. Rochette, S. Ayotte, and L. A. Rusch "Analysis of the Spectral Efficiency of Frequency-Encoded OCDMA System with Incoherent Sources," *J. Lightwave Technol.*, vol. 23, no. 4, pp. 1610–1619, Apr. 2005.

[10] C. H. Lin, J. Wu, H. W. Tsao, and C. L. Yang "Spectral Amplitude-Coding Optical CDMA System Using Mach–Zehnder Interferometers," *J. Lightwave Technol.*, vol. 23, no. 4, pp. 1543–1555, Apr. 2005.

[11] B. Moslehi, "Analysis of Optical Phase Noise in Fiber Optic Systems Employing a Laser Source with Arbitrary Coherence Time," *J. Lightwave Technol.*, vol. 4, no. 9, pp. 1334–1351, Sep. 1986.